\documentclass[twocolumn,twoside]{IEEEtran}
\usepackage{amsmath,amssymb,amsthm,epsfig,color,subfigure,empheq,graphicx,pgfplots}
\usepackage{enumerate,url,algorithm,wasysym,epstopdf,balance,enumerate}

\newcommand \bb{\mathbf{b}}
\newcommand \be{\mathbf{e}}
\newcommand \bm{\mathbf{m}}
\newcommand \bn{\mathbf{n}}

\newcommand \bv{\mathbf{v}}
\newcommand \bw{\mathbf{w}}
\newcommand \bx{\mathbf{x}}
\newcommand \by{\mathbf{y}}

\newcommand \bA{\mathbf{A}}
\newcommand \bC{\mathbf{C}}

\newcommand \bF{\mathbf{F}}
\newcommand \bI{\mathbf{I}}
\newcommand \bJ{\mathbf{J}}

\newcommand \bN{\mathbf{N}}
\newcommand \bQ{\mathbf{Q}}

\newcommand \bU{\mathbf{U}}
\newcommand \bV{\mathbf{V}}
\newcommand \bX{\mathbf{X}}

\newcommand \hbx{\hat{\mathbf{x}}}

\newcommand \0{\mathbf{0}}
\newcommand \bdelta{\boldsymbol{\delta}}
\newcommand \bmu{\boldsymbol{\mu}}
\newcommand \bsigma{\boldsymbol{\sigma}}
\newcommand \bxi{\boldsymbol{\xi}}

\newcommand \bbeta{\boldsymbol{\beta}}
\newcommand \bLambda{\boldsymbol{\Lambda}}

\newcommand \bSigma{\boldsymbol{\Sigma}}

\newtheorem{assumption}{Assumption}

\DeclareMathOperator{\diag}{dg}

\DeclareMathOperator{\trace}{Tr}

\newtheorem{proposition}{Proposition}
\newtheorem{lemma}{Lemma}

\newtheorem{theorem}{Theorem}

\newtheorem{remark}{Remark}

\begin{document}


\title{Online State Estimation for Time-Varying Systems}
	
\author{Guido Cavraro, Emiliano Dall'Anese, Joshua Comden, Andrey Bernstein



\thanks{G. Cavraro, J. Comden, and A. Bernstein are with the National Renewable Energy Laboratory (NREL), Golden CO, USA (Email: {\tt\small \{guido.cavraro, joshua.comden, andrey.bernstein\}@nrel.gov})
E. Dall'Anese is with the University of Colorado Boulder, Boulder, CO, USA (Email: {\tt\small emiliano.dallanes@colorado.edu}). The work of E. Dall'Anese was supported by NREL award APUP UGA-0-41026-109. }

}	


\maketitle

\begin{abstract}
The paper investigates the problem of estimating the state of a time-varying system with a linear measurement model; in particular, the paper considers the case where the number of measurements available can be smaller than the number of states. In lieu of a batch linear least-squares (LS) approach -- well-suited for static networks, where a sufficient number of measurements could be collected to obtain a full-rank design matrix -- the paper proposes an online algorithm to estimate the possibly time-varying state by processing measurements as and when available. The design of the algorithm hinges on a generalized LS cost augmented with a proximal-point-type regularization. With the solution of the regularized LS problem available in closed-form, the online algorithm is written as a linear dynamical system where the state is updated based on the previous estimate and based on the new available measurements. Conditions under which the algorithmic steps are in fact a contractive mapping are shown, and bounds on the estimation error are derived for different noise models. Numerical simulations are provided to corroborate the analytical findings. 

\end{abstract}


\section{Introduction}\label{sec:intro}

State estimation plays a crucial role in large-scale engineering systems --  including traffic, energy, and communication networks -- because it is essential for monitoring purposes and to support underlying control and optimization tasks. For instance, state estimation in power systems pertains to the reconstruction of voltage profiles given a set of sparse measurements~\cite{Kekatos}; in traffic networks, traffic flows and vehicle densities in highways and roads are monitored and used for congestion control~\cite{CASCETTA1984289}.
Estimating the state of a network may be challenging, since oftentimes key quantities are not directly accessible or are not constantly measured. For instance, event-triggered communication mechanisms are introduced to reduce unwanted network traffic and energy consumption~\cite{Zong,Chen}.

\begin{figure}[t]
\centering
\includegraphics[width=0.4\textwidth]{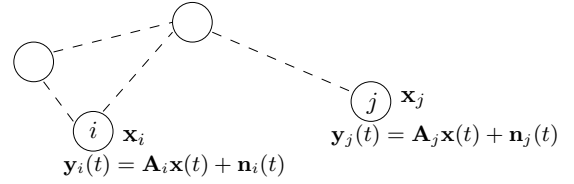}
\caption{Networked system of agents. Each agent is able to measure locally the noisy version of a linear function of the whole system state.}
\label{fig:sys_network}
\end{figure}

In this paper, we consider the memoryless model\footnote{\emph{Notation} lower- (upper-) case boldface letters denote column vectors (matrices). Calligraphic symbols are reserved for sets. Symbol $^{\top}$ stands for transposition. Vectors $\mathbf{0}$ and $\mathbf{1}$ are the all-zero and all-one vectors, while $\be_m$ is the $m$-th canonical vector. Symbol $\|\mathbf{x}\|$ and $\|\mathbf{X}\|$ denote the $2$-norm of the vector $\mathbf{x}$ and of the matrix $\bX$, respectively; symbol $\|\mathbf{X}\|_F$ denotes the Frobenius norm of $\bX$, while $\|\bx\|_{\bQ}  = \bx^\top \bQ \bx$ for a positive definite matrix $\bQ$.
The diagonal matrix having the elements of the finite set $\{x_i\} = \{x_1, x_2, \dots \}$ on its diagonal is denoted as $\diag(\{x_i\})$.
Given a matrix $\bA$, its kernel, namely the set of all vectors $\bx$ such that $\bA \bx = \0$, is denoted as $\ker \bA$. The expectation operator is defined as $\mathbb E[\cdot]$.
The Kronecker product of the vectors $\bx$ and $\bx'$ is $\bx \otimes \bx'$, while $\text{vec} (\bX)$ is the vectorization of the matrix $\bX$. Finally, given a sequence of matrices $\{\bX(t)\}_{t=1}^T$, we have that $\prod_{t=1}^T \bX(t) = \bX(T) \bX(T-1) \dots \bX(1)$.
} 
\begin{equation}
\by(t) = \bA(t) \bx(t) + \bn(t)
\label{eq:sys_model}
\end{equation}
where $\by(t)$ is a vector of available measurements at time $t$, the system state is represented by the vector $\bx(t)$, $\bn(t)$ is the vector of noise, and $\bA(t)$ is a possibly time-varying regression matrix.
The general model~\eqref{eq:sys_model} is representative of state estimation tasks in several applications of interest, e.g., in wireless sensor networks or power systems~\cite{SCHENATO20111878,Cav_GS19}; in can also represent the output measurement equation of given dynamical systems. The main motivation behind the time-variability of $\bA(t)$ is the following.
First, the structure of the underlying system could be time-varying. For instance, the topology of a power network could be modified in order to optimize the network performance. In this case, different network topologies would be associated with different regression matrices.
Second, a system could have sensors that report measurements infrequently and at different times~\cite{HU2017145}.
To elaborate further, consider the network of agents depicted in Figure~\ref{fig:sys_network}, in which each agent $k$ measures the quantity 
\begin{equation}
    \by_k(t) = \bA_k(t) \bx(t) + \bn_k(t)
    \label{eq:y_i}
\end{equation}
but only a subset of the agents reports the measurements. In this scenario, $\by(t)$ and $\bA(t)$ are obtained by stacking the $\by_i(t)$'s and the $\bA_i(t)$'s associated with agents that reported their measurement at time $t$. For instance, assume that only node $i$ and node $j$ in Figure~\ref{fig:sys_network} send measurements at time~$t$. Then, we have $\by(t) = \begin{bmatrix} \by_i(t)^\top & \by_j(t)^\top\end{bmatrix}^\top$, and $\bA(t) = \begin{bmatrix} \bA_i(t)^\top & \bA_j(t)^\top\end{bmatrix}^\top$. 
Notable examples of such systems are power networks in which sensors, like smart meters, do not provide synchronized measurements, i.e., the measurements are not taken at the same time~\cite{Cav_GS19,Alimardani}; or battery-powered sensor networks where measurements are parsimoniously collected and transmitted to strike a balance between estimation accuracy and energy consumption~\cite{Alippi}.

When a sufficient number of measurements can be collected before the state $\bx(t)$ changes and the regression matrix is full rank, state estimation is classically performed via least squares methods~\cite{Swerling}; to handle underdetermined systems, pertinent regularized counterparts, maximum likelihood or Bayesan approaches can be pursued~\cite{kay1993fundamentals}.
In \emph{dynamic settings} where the system state evolves in time~\cite{popkov2005gradient,SimonettoGlobalsip2014} and streams of measurements are received asynchronously~\cite{dall2020optimization}, the time-variability of the state might be such that a sufficient number of measurements to obtain a unique state estimate cannot be collected. The fusion of data from multi-rate asynchronous sensors with measurements randomly missing is studied in~\cite{Yan2010}. Missing data and delays are likely to occur in asynchronous multi-sensor systems. Algorithms suited for this scenario have been proposed in~\cite{MAHMOUD201215,Matveev}.
It is worth pointing out that Kalman filtering can be used to perform state estimation in dynamic systems described by a state space model~\cite{kay1993fundamentals}. Customized extensions of the Kalman filter have been tailored to handle systems where measurements are both taken by sensors and collected by the system operators at different times. In~\cite{Sinopoli}, the Kalman filter was generalized for the case in which the arrival of observation is modeled as a random process that depends on the communication channel features.

This paper considers a setting in which
    the number of available measurements at every time step is much smaller than the number of state variables,
    and 
    a meaningful (deterministic or stochastic) state space description is not available and hence traditional least squares estimators or Kalman filter-based approaches cannot be pursued.
We propose an online state estimator (OSE) that, at each time step, solves a strongly convex optimization problem. Its cost is the sum of a weighted least squares term, that captures the available measurements data, and a regularization term, that introduces ``memory'' on the estimate by feeding back the previous-step estimation. This momentum term ensures a consistent and accurate estimate under low-observability conditions.

We show that the state estimate follows a dynamic linear system, with the measurements as an input. We then analyze the performance of this system under bounded deterministic and zero-mean stochastic noise assumptions.
These two cases are both meaningful: in the first, the noise can be interpreted as a bounded modeling error, in the second, as measurement noise. 
The main contribution of this paper is to show that the OSE can track the true system state up to a bounded error (an error with bounded mean and variance) when a bounded (stochastic) measurement noise is introduced.
\section{The State Estimator}
\label{sec:prob_form}

Consider a discrete time system whose state at time $t = 0,1,\dots$ is described by the vector $\bx(t) \in \mathbb{R}^N$ and whose output $\by(t)$ is modeled by~\eqref{eq:sys_model},
%
%
where $\by(t), \bn(t) \in \mathbb{R}^{M_t}, \bA(t) \in \mathbb{R}^{M_t \times N}$ and where $M_t$ is allowed to vary in time. In the following, the vector $\bn$ will be referred to as \emph{noise vector} since it has a straightforward interpretation as the measurement noise affecting the system output $\by$; nevertheless, $\bn$ can also be used to describe model uncertainty\footnote{Potentially, measurement vectors could be available at times $t = t_1, t_2, \dots$ not equally spaced. However, to keep the notation simple and without loss of generality, in the following we will assume that measurements are produced at times $t = 1,2, \dots$}.
The system state $\bx$ is assumed to be time-varying and the state variation at time $t$ is denoted as
\begin{equation}
    \bdelta(t) := \bx(t) -  \bx(t-1).
    \label{eq:x_var}
\end{equation}

A model describing how $\bx$ changes in time, e.g., a state space model, is not available. Rather, mild information on the state variation is assumed to be known.
Precisely, for every $t$, there exists a real non-negative scalar $\Delta_{x}(t)$ such that
\begin{equation}
\| \bdelta(t)\| \leq \Delta_{x}(t).
\label{eq:x_bounded}
\end{equation}
Further, let $\Delta_{x} := \sup\{\Delta_{x}(t)\}$, and suppose $\Delta_{x} < \infty$. 

This paper proposes an algorithm that provides an estimate $\hbx$ of the system state $\bx$ given the system output $\by$ and the sequence of model matrices $\{\bA(t)\}_{t \geq 1}$.
A straightforward way to obtain $\hbx(t), t \geq 1$ would be solving the Weighted Least Square (WLS) problem
\begin{equation}
\arg \min_\bw \| \by(t) - \bA(t) \bw \|_{\bQ_t^{-1}}^2
\label{eq:WLS}
\end{equation}
where $\bQ_t \in \mathbb{R}^{M_t \times M_t}$ is a positive definite matrix. Problem~\eqref{eq:WLS} has a unique solution only if the number of measurements available is greater or equal to the number of system's states, namely, $M_t \geq N$.
Otherwise, \eqref{eq:WLS} is not strictly convex and has infinitely many solutions.
Since the focus of this paper is on systems in which possibly $M_t \ll N$, the WLS approach can not be pursued. Rather, we propose to compute the state estimate by solving the following time-varying \emph{regularized WLS problem} for $t = 1, 2, \ldots $ and given an initial $\hbx(0)$
\begin{equation}
\hbx(t) = \arg \min_\bw \| \by(t) - \bA(t) \bw \|_{\bQ_t^{-1}}^2 + \gamma \| \bw - \hat{\bx}(t-1)\|^2.
\label{eq:DASE_WLS}
\end{equation}
The second term in~\eqref{eq:DASE_WLS} acts as a regularizer which penalizes the Euclidean distance of the new estimate from the older one and makes \eqref{eq:DASE_WLS} a strongly convex problem having a unique solution. The real scalar $\gamma > 0$ will be referred to as the \emph{inertia parameter}.
The smaller $\gamma$ is, the further the new estimate $\hbx(t)$ is allowed to be from $\hbx(t-1)$.
The estimate $\hbx(t)$ admits the closed form
\begin{equation}
\hbx(t) = \bLambda(t) \hbx(t-1) + \frac 1 \gamma \bLambda(t) \bA(t)^\top \bQ^{-1}_t \by(t)
\label{eq:DASE_update}
\end{equation}
where
\begin{equation}
\bLambda(t) := \gamma (\bA(t)^\top \bQ^{-1}_t \bA(t) + \gamma \bI)^{-1}.
\label{eq:Lambda}
\end{equation}
That is, the new estimate $\hbx(t)$ can be computed recursively given the previous estimate $\hbx(t-1)$, the new measurement $\by(t)$, and the new $\bA(t)$.
Equation~\eqref{eq:DASE_update} represents the sought online asynchronous state estimator.
The inverse on the right hand side of \eqref{eq:Lambda} always exists and $\bLambda(t) \in \mathbb{R}^{N \times N}$ is a symmetric positive definite matrix for every $t$. 

Next, the matrix $\bLambda(t)$ is characterized. To this aim, consider the matrix $\bJ(t) = \bA(t)^\top \bQ_t^{-1} \bA(t)$, which is a positive semi-definite $N \times N$ matrix and admits the following decomposition
\begin{equation}
\bJ(t) = 
\begin{bmatrix}
\bU(t)&\bV(t)
\end{bmatrix}
\begin{bmatrix}
\diag(\{\lambda_i(t)\})& \0 \\
\0 & \0
\end{bmatrix}
\begin{bmatrix}
\bU^\top(t) \\ \bV^\top(t)
\end{bmatrix}
\label{eq:svd_J}
\end{equation}
where $\lambda_i(t)$ is the $i$-th non zero eigenvalue of $\bJ(t)$ with $0 < \lambda_1(t) \leq \lambda_2(t) \leq \dots$. The matrices $\bV(t) \in \mathbb{R}^{N \times K_t}$ and $\bU(t)\in \mathbb{R}^{N \times I_t}$ collect the eigenvectors of $\bJ(t)$ associated with zero eigenvalues and non-zero eigenvalues, respectively.
Hence, $\bV(t)$ spans $\ker \bJ(t)$, which is a space of dimension $K_t$; $\bU(t)$ spans the image of $\bJ(t)$, which is a space of dimension $I_t = N - K_t$. Notably, $\ker \bJ(t)$ coincides with $\ker \bA(t)$, as shown in the next result.
\begin{lemma}
A vector $\bv \in \mathbb{R}^N$ is in the kernel of $\bJ(t)$, $\bv \in \ker \bJ(t)$, if and only if $\bv$ is in the kernel of $\bA(t)$, $\bv \in \ker \bA(t)$. 
\end{lemma}
\begin{IEEEproof}
If $\bv \in \ker \bA(t)$, trivially $\bv \in \ker \bJ(t)$. Now assume $\bv \in \ker \bJ(t)$. Then, $\bA(t) \bv = 0$ because
$
  \|\bA(t) \bv\|_{\bQ_t^{-1}} =  \bv^\top  \bA(t)^\top \bQ_t^{-1} \bA(t) \bv = \bv^\top \0 = 0.
$
\end{IEEEproof}
Using equation~\eqref{eq:svd_J}, being $\bLambda(t) := \gamma (\bJ(t) + \gamma \bI)^{-1}$, we find that $\bLambda(t)$ is the symmetric positive definite matrix
\begin{equation}
\bLambda(t) = 
\begin{bmatrix}
\bU(t)&\bV(t)
\end{bmatrix}
\begin{bmatrix}
\diag \left (\left\{\frac{\gamma}{\gamma + \lambda_i(t)}\right\} \right)& \0 \\
\0 & \bI
\end{bmatrix}
\begin{bmatrix}
\bU^\top(t) \\ \bV^\top(t)
\end{bmatrix}.
\label{eq:svd_Lam}
\end{equation}
Hence, matrices $\bJ(t)$ and $\bLambda(t)$ share the same eigenvectors and the spectrum of $\bLambda(t)$ is given by 
\begin{equation}
\text{eig }\bLambda(t) = \left\{1, \frac{\gamma}{\gamma + \lambda_1(t)}, \dots,  \frac{\gamma}{\gamma + \lambda_{I_t}(t)} \right\}
\label{eq:spect_Lambda}
\end{equation}
where 1 is an eigenvalue with multiplicity $K_t$ and $1 \geq \frac{\gamma}{\gamma + \lambda_1(t)} \geq \frac{\gamma}{\gamma + \lambda_2(t)} \geq \dots  $.
Moreover, $\|\bLambda(t)\| \leq 1$ and
\begin{equation}
\bx \mapsto \bLambda(t) \bx
\label{eq:xtoLx}
\end{equation}
is a non-expansive operator in general. Finally, let 
$\bar \lambda$ denote the smallest non-zero eigenvalue 
for all $\bJ(t)$:
$$ \bar \lambda := \min \{ \lambda_1(t), t \geq 1\}.$$ 

\begin{figure}[t]
\centering
\includegraphics[width=0.3\textwidth]{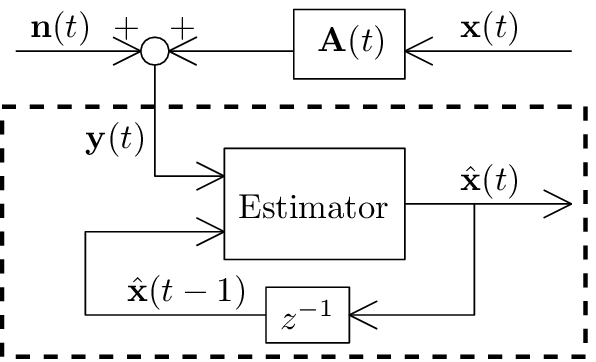}
\caption{Block scheme of the dynamical system described by equation \eqref{eq:DASE_update}.}
\label{fig:sys_block_x}
\end{figure}

Equations \eqref{eq:sys_model} and \eqref{eq:DASE_update} constitute a linear dynamical system, whose block scheme is reported in Figure~\ref{fig:sys_block_x}. Furthermore, heed that equation \eqref{eq:DASE_update} is essentially a classic closed-loop system.

Before proceeding, the next result can be used to provide a familiar interpretation for equation~\eqref{eq:DASE_update}. 

\begin{lemma}
Consider the matrix $\bLambda(t)$. It holds
\begin{equation}
\Big(\bI - \bLambda(t)\Big) \bx = \frac 1 \gamma \bLambda(t) \bA(t)^\top \bQ^{-1} \bA(t) \bx.
\label{eq:lemma1}    
\end{equation}
\label{lem:lemma1}  
\end{lemma}

\begin{IEEEproof}
Since $\bLambda(t)$ is a positive definite matrix, and its inverse always exists, it holds
\begin{align*}
\bx & = \bLambda(t) \bLambda(t)^{-1} \bx = \frac 1 \gamma \bLambda(t) \big(\bA(t)^\top \bQ^{-1} \bA(t) + \gamma \bI \big) \bx \\
    & = \bLambda(t) \bx + \frac 1 \gamma \bLambda(t) \bA(t)^\top \bQ^{-1} \bA(t) \bx 
\end{align*}
from which equation~\eqref{eq:lemma1} follows.
\end{IEEEproof}

Reminding that we defined $f_t(\bw) = \| \by(t) - \bA(t) \bw \|_{\bQ_t^{-1}}^2$, we have that
$$ \nabla f_t(\hbx(t-1)) = \bA(t)^\top \bQ_t^{-1} \big(\by(t) - \bA(t)\bw \big);$$
equation~\eqref{eq:lemma1} can be used to rewrite~\eqref{eq:DASE_update} as
\begin{align*}
\hbx(t) & = \hbx(t-1) - \frac 1 \gamma \bLambda(t) \bA(t)^\top \bQ^{-1}_t \left(\by(t) - \bA(t)\bw \right) \\
 & = \hbx(t-1) - \frac 1 \gamma \bLambda(t) \nabla f_t(\hbx(t-1)).
\label{eq:DASE_update_GD}
\end{align*}
Being $\bLambda(t)$ a positive definite matrix, $\bLambda(t) \nabla f_t(\hbx(t-1))$ is a descent direction for the function $f_t(\bw)$, i.e., $\hbx(t)$ is computed, for every $t$, via a Newton-like descent of $f_t(\bw)$.

\begin{remark}
The OSE~\eqref{eq:DASE_update} can be  adapted for the more general case in which the system output is a noisy version of a linear affine function of the system state, namely,
\begin{equation*}
    \by(t) = \bA(t) \bx(t) + \bb(t) + \bn(t),
\end{equation*}
where $\bb(t) \in \mathbb R^{M_t}$. In fact, it is enough to define the variable $\tilde \by(t) = \by(t) - \bb(t)$ and then compute the state estimate via
\begin{equation*}
\hbx(t) = \bLambda(t) \hbx(t-1) + \frac 1 \gamma \bLambda(t) \bA(t)^\top \bQ^{-1}_t \tilde \by(t)
\end{equation*}
We consider systems like~\eqref{eq:sys_model} to
reduce needed notations.
\end{remark}

\begin{remark} \label{rem:ppm}
From the optimization perspective, the mathematical formulation of~\eqref{eq:DASE_WLS} is the one of the proximal point method (PPM)~\cite{parikh2014,rockafellar1976monotone}. The PPM is an algorithm aiming at minimizing a function $f(\bx)$ by iteratively solving the problem~\cite{parikh2014}
\begin{equation}
\hbx(t) = \arg \min_\bw f(\bw) + \frac{1}{2 \lambda} \|\bw - \hbx(t-1)\|^2.
\label{eq:ppm}
\end{equation}
After denoting the first term of the cost 
in~\eqref{eq:DASE_WLS} as $f_t(\bw) = \| \by(t) - \bA(t) \bw \|_{\bQ_t^{-1}}^2$
we can rewrite~\eqref{eq:DASE_WLS} as
\begin{equation}
\hbx(t) = \arg \min_\bw f_t(\bw) + \gamma \|\bw - \hbx(t-1)\|^2,
\label{eq:dse_cmp}
\end{equation}
which is a time-varying PPM for the sequence of functions $\{f_t\}$.
The main difference is that, whereas the PPM is used to find iteratively a solution of a static optimization problem, we are considering the case in which the optimization problem changes at every iteration and the goal is to \emph{track} its solution 
which represents the true system state.

\end{remark}

\section{The estimator's performance}
\label{sec:es_perf}

Define the estimation error $\bxi$, namely, the difference between the state estimate and the true state for $t \geq 1$, as
\begin{equation}
\bxi(t) = \hbx(t) - \bx(t). 
\label{eq:est_err}
\end{equation}
Like $\bx$, the error $\bxi$ has a closed form expression whose derivation is possible thanks to the following result.

To obtain the estimation error closed form expression, substitute~\eqref{eq:sys_model} into~\eqref{eq:DASE_update},
%
%
use equation~\eqref{eq:lemma1} and equation~\eqref{eq:est_err}
%
%
%
\begin{equation}
\bxi(t) = \bLambda(t) \bxi(t-1) - \bLambda(t) \bdelta(t) + \frac 1 \gamma \bLambda(t) \bA(t)^\top \bQ^{-1}_t \bn(t).
\label{eq:est_err_update}
\end{equation}
By iteratively applying \eqref{eq:est_err_update}, we can find the expression of $\bxi(T)$, for every $T\geq 1$, namely

\begin{small}
\begin{equation}
    \bxi(T) = \prod_{t=1}^{T} \bLambda(t) \bxi(0) + \sum_{t=1}^T \prod_{k=t}^T \bLambda(k) \left(\frac 1 \gamma \bA(t)^\top \bQ_t^{-1} \bn(t) -\bdelta(t) \right)
    \label{eq:xi(T)}
\end{equation}
\end{small}

\noindent where $\bxi(0)$ is the initial estimation error.

We next analyze the OSE performance under two conditions on the noise. 
In the first case, $\bn$ is assumed to be a vector whose norm is bounded. This corresponds to scenarios in which $\bn$ represents a modeling error that is known to be finite. In the second case, $\bn(t)$ is assumed to be a stochastic vector with a certain mean and variance. This case can describe scenarios in which $\bn$ represents the measurement error.
%
%
\begin{figure}[t]
\centering
\includegraphics[width=0.4\textwidth]{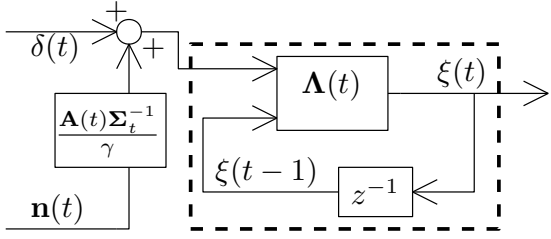}
\caption{Block scheme of the dynamical system described by equation \eqref{eq:est_err_update}.}
\label{fig:sys_block}
\vspace{-.4cm}
\end{figure}

Next, the estimation error is characterized.
We make the following assumption regarding the model matrices $\{\bA_t\}$.
\begin{assumption}
There exists a constant $\tau > 0$ such that
\begin{equation}
\bigcap_{k = 0}^{\tau-1} \ker \bA(t + k) = \{\0\}, \quad t \geq 1.
\label{eq:kernel}
\end{equation}
\label{ass:tau}
\end{assumption}

Roughly speaking, Assumption \ref{ass:tau} means that every $\tau$ time steps, the system is  fully observable; this will be quantified precisely in Propositions \ref{prop:asymp_stable} and \ref{prop:sigma_asymp_stable} below.


\paragraph{Bounded Noise}
\label{sub:bnd_noise}
Next, the case in which $\bn$ is a bounded unknown vector will be considered.
\begin{assumption}
The noise vector $\bn$ is bounded, i.e., there exists a real non-negative scalar $\Delta_{n}(t)$ such that
\begin{equation}
\| \bn(t)\| \leq \Delta_{n}(t).
\label{eq:n_bounded}
\end{equation}
Further, let $\Delta_{n} := \sup\{\Delta_{n}(t)\}$, and suppose $\Delta_n < \infty$.
%
%
\label{ass:n_bounded}
\end{assumption}
The results reported hereafter are proved in the Appendix. Assumption~\ref{ass:tau} has as a direct consequence the next Proposition, which will be used next to prove the main result.
\begin{proposition}
Consider the system described by 
\begin{equation}
    \bxi(t) = \bLambda(t) \bxi(t-1).
\label{eq:din_sys}
\end{equation}
and define
\begin{equation*}
\psi = \max_{t} \left\{\frac{\gamma}{\gamma + \lambda_1(t)} \right \}.
\end{equation*}
Then, it holds that
\begin{equation}
    \left \| \prod_{k = 0}^{\tau-1} \bLambda(t + k) \right \| \leq \psi < 1 
    \label{eq:norm_prod_str_low1}
\end{equation}
and the system~\eqref{eq:din_sys} is asymptotically stable
\begin{equation*}
\lim_{T \rightarrow \infty} \bxi(T) = \prod_{t=1}^T \bLambda(t) \bxi(0) = \0.
\end{equation*}
\label{prop:asymp_stable}
\end{proposition}

Note that Proposition~\ref{prop:asymp_stable} implies that the operator
\begin{equation*}
\bx \mapsto \bLambda(t+\tau-1) \bLambda(t+\tau-2) \dots \bLambda(t)\bx
\end{equation*}
is a contraction even if the map in~\eqref{eq:xtoLx} is not contractive.
The estimation error meets the next property.
\begin{theorem}
Let Assumptions~\ref{ass:tau} and~\ref{ass:n_bounded} hold. Define $c(t) :=  \|\bA(t)^\top \bQ_t^{-1}\|$. The estimation error at time $T$ is upper bounded by
\begin{equation}
\|\bxi(T)\| \leq  \psi^{\left \lfloor{\frac{T}{\tau}}\right \rfloor} \|\bxi(0)\| + \sum_{t = 1}^T \psi^{ \left \lfloor{\frac{T+1-t}{\tau}}\right \rfloor} \Big(\Delta_{x}(t) + \frac{c(t)}{\gamma} \Delta_{n}(t) \Big).
\label{eq:_xi_bnd_bnd_T}
\end{equation}
Moreover, define the constant $c := \sup_t \|\bA(t)^\top \bQ_t^{-1}\|$.
The estimation error is asymptotically upper-bounded, i.e., 
    \begin{equation}
        \limsup_{t \rightarrow \infty} \|\bxi(t)\| \leq  \tau \Big(\Delta_x + \frac 1 \gamma c \Delta_n\Big) \Big(1+ \frac{\gamma}{\bar \lambda}\Big).
        \label{eq:_xi_bnd_bnd}
    \end{equation}
Finally, the error upper bound in~\eqref{eq:_xi_bnd_bnd} is minimized by
\begin{equation}
\gamma^* = \sqrt{ \frac{c \bar \lambda \Delta_n}{\Delta_x}}.
\label{eq:gamm_star_bnd}
\end{equation}
\label{thm:err_bounded}
\end{theorem}


\paragraph{Stochastic Noise}

Here, the case in which $\bn$ is a random vector will be considered.

\begin{assumption}
The noise vector $\bn$ is an i.i.d. random vector with zero-mean and finite positive definite covariance $\bN_t \in \mathbb{R}^{M_t \times M_t}$, 
%
$\mathbb E [ \bn(t) ] = \0, \quad \mathbb E [ \bn(t) \bn(t)^\top] = \bN_t$.
%
\label{ass:n_stoch}
\end{assumption}

In this case, a standard choice is to set $\bQ_t = \bN_t$.
Denote as $\bmu(t) := \mathbb E[\bxi(t)]$ and $\bSigma(t) := \mathbb E[(\bxi(t)-\bmu(t))(\bxi(t)-\bmu(t))^\top]$ the mean and the covariance of the estimation error at time $t$. Given Assumption~\ref{ass:n_stoch} and by applying the expectation operator to~\eqref{eq:xi(T)}, at every time $T \geq 1$ we have that
\begin{equation}
\bmu(T)  = \prod_{t=1}^{T} \bLambda(t) \bxi(0) - \sum_{t=1}^T \prod_{k=t}^T \bLambda(k)  \bdelta(t).
\label{eq:mu(T)}
\end{equation}
Equation~\eqref{eq:mu(T)} can be used to compute also the error covariance at time $T$:
\begin{align}
& \bSigma(T)  = \mathbb E[(\bxi(T)-\bmu(T))(\bxi(T)-\bmu(T))^\top] \notag \\
& = \frac{1}{\gamma^2} \mathbb E \Bigg[ \left( \sum_{t=1}^T \left(\prod_{k=t}^T \bLambda(k)\right)  \bA(t)^\top \bN_t^{-1} \bn(t) \right) 
\notag \\
& \qquad \qquad \left( \sum_{t=1}^T \left(\prod_{k=t}^T \bLambda(k)\right)  \bA(t)^\top \bN_t^{-1} \bn(t) \right)^\top \Bigg] \notag \\
& = \sum_{t=1}^T \left(\prod_{k=t}^T \bLambda(k) \right)  \frac{\bA(t)^\top \bN_t^{-1}  \bA(t)}{\gamma^2} \left(\prod_{k=t}^T \bLambda(k) \right)^\top.
\label{eq:Sigma(T)}
\end{align}

Similar computations can be used to find $\bSigma(T+1)$

\begin{small}
\begin{align}
&\bSigma (T+1) 
= \sum_{t=1}^{T} \left(\prod_{k=t}^{T+1} \bLambda(k) \right)  \frac{\bA(t)^\top \bN_t^{-1}  \bA(t)}{\gamma^2}  \left(\prod_{k=t}^{T+1} \bLambda(k) \right)^\top \notag \\
&  \quad +  \bLambda(T+1) \frac{\bA(T+1)^\top \bN_t^{-1} \bA(T+1)}{\gamma^2} \bLambda(T+1)^\top.
\label{eq:Sigma(T+1)}
\end{align}
\end{small}

Comparing equations~\eqref{eq:Sigma(T)} and~\eqref{eq:Sigma(T+1)}, it can be shown that the error covariance obeys the linear system
\begin{align}
\bSigma &(t+1)= \bLambda(t+1) \bSigma(t)  \bLambda(t+1)^\top \notag \\
& \quad +  \bLambda(t+1) \frac{\bA(t+1)^\top \bN_t^{-1}\bA(t+1)}{\gamma^2} \bLambda(t+1)^\top.
\label{eq:Sigma_lin_sys}
\end{align}

To conveniently study the estimation error variance, introduce the vector $\bsigma(t) := \text{vec }( \bSigma(t))$, $\bsigma(t) \in \mathbb{R}^{N^2}$. By exploiting the well known properties of the Kronecker product, the evolution of $\bsigma$ can be expressed as
\begin{equation}
\bsigma(t) = \bF(t) \bsigma(t-1) + \frac{1}{\gamma^2} \bF(t) \bC(t) \bm(t)
\label{eq:sigma_it_upd}
\end{equation}
where $\bF(t) := \bLambda(t) \otimes \bLambda(t)$, $\bC(t) := \bA(t)^\top  \otimes  \bA^\top(t)$, and $\bm(t) := \text{vec }(\bN_t^{-1})$.
Iterating equation~\eqref{eq:sigma_it_upd} yields, for $T\geq 1$,
\begin{equation}
\bsigma(T) = \prod_{t=1}^T \bF(t) \bsigma(0) + \frac{1}{\gamma^2} \sum_{t=1}^{T} \prod_{k=t}^T \bF(k) \bC(t) \bm(t) 
\label{eq:sigma(T)}
\end{equation}
The results reported hereafter are proved in the Appendix. Firstly, we provide a direct consequence of Assumption~\ref{ass:tau}.
\begin{proposition}
Consider the system described by 
\begin{equation}
    \bsigma(t) = \bF(t) \bsigma(t-1).
\label{eq:F_din_sys}
\end{equation}
It holds that
\begin{equation}
    \left \| \prod_{k = 0}^{\tau-1} \bF(t + k) \right \| \leq \psi < 1 
    \label{eq:F_norm_prod_str_low1}
\end{equation}
and the system~\eqref{eq:din_sys} is asymptotically stable
\begin{equation*}
\lim_{T \rightarrow \infty} \bsigma(T) = \prod_{t=1}^T \bF(t) \bsigma(0) = \0.
\end{equation*}
\label{prop:sigma_asymp_stable}
\end{proposition}

Proposition~\ref{prop:sigma_asymp_stable} is used to prove the next main result.

\begin{theorem}
Let Assumptions~\ref{ass:tau} and ~\ref{ass:n_stoch} hold, and define $C(t) :=  \|\bA^\top(t) \otimes \bA^{\top}(t)\|_F$ and $m(t) :=  \|\bN_t^{-1}\|_F$.
\item The error mean and error variance at time $T$ are such that
\begin{align}
&\|\bmu(T)\| \leq  \psi^{\left \lfloor{\frac{T}{\tau}}\right \rfloor} \|\bxi(0)\| + \sum_{t = 1}^T \psi^{ \left \lfloor{\frac{T+1-t}{\tau}}\right \rfloor} \Delta_{x}(t) \label{eq:mu_stoch_T}\\
&\|\bSigma(T)\|_F \leq \psi^{\left \lfloor{\frac{T}{\tau}}\right \rfloor}\|\bSigma(0)\|_F + \sum_{t = 1}^T \psi^{ \left \lfloor{\frac{T+1-t}{\tau}}\right \rfloor} C(t) m(t).
\label{eq:sigma_stoch_T}
\end{align}
Moreover, define $C := \sup_t \{ C(t) \}$ and $m := \sup_t \{ m(t) \}$. The error mean, the error variance, and the average distance between the estimate $\hbx$ and the true state $\bx$, namely $ \sqrt{\mathbb E[(\hbx - \bx)^\top (\hbx - \bx) ]} = \sqrt{\mathbb E[\|\bxi\|^2}]$ is asymptotically upper-bounded 
    \begin{align}
        &\limsup_{t \rightarrow \infty} \|\bmu(t)\| \leq \tau \Delta_x \left(1 + \frac{\gamma}{\bar \lambda} \right) \label{eq:mu_stoch}\\
        &\limsup_{t \rightarrow \infty} \|\bSigma(t)\|_F \leq \frac{ \tau C m}{\gamma^2} \left(1 + \frac{\gamma}{\bar \lambda} \right)
        \label{eq:sigma_stoch}\\
        &\limsup_{t \rightarrow \infty}  \sqrt{\mathbb E \|\bxi(t)\|^2} \leq \tau  \sqrt{\frac{C^2 m^2}{\gamma^4} + \Delta_x^2} \left(1 + \frac{\gamma}{\bar \lambda} \right).
        \label{eq:ave_xi_stoch}
    \end{align}
\label{thm:err_stoch}
\end{theorem}
%
%
%
%
%
%
%
%

\begin{remark}
Theorem~\ref{thm:err_bounded} and Theorem~\ref{thm:err_stoch} have been derived essentially by studying the bounded input-bounded output (BIBO) stability properties of the systems~\eqref{eq:est_err_update}, \eqref{eq:mu(T)}, and~\eqref{eq:sigma_it_upd}; see the region within the dashed rectangle in Figures~\ref{fig:sys_block}. In~\cite{Michaletzky}, the BIBO stability is proved for linear switching systems which are uniformly exponentially stable. These are systems for which, given an initial condition $\bx(0)$ and when the input is identically zero, there exists a $\lambda < 1$ and a $c < 1$ such that the norm of the state $\bx$ can be bounded as
$$\| \bx(t) \| \leq c \lambda^t \| \bx(0) \|$$
for any $t\geq 1$ and for any switching path. Unfortunately, this is not the case for the systems~\eqref{eq:est_err_update}, \eqref{eq:mu(T)}, and~\eqref{eq:sigma_it_upd}, for which a similar property holds but only once every $\tau$ time steps.
\end{remark}

\begin{remark}
Heed that the estimation error is finite for any inertia parameter meeting the condition $\gamma < \infty$. That is, for any finite choice of $\gamma$, the estimation errors upper bounded by~\eqref{eq:_xi_bnd_bnd} and~\eqref{eq:ave_xi_stoch} do not diverge.
\end{remark}


\section{Numerical Results}
\label{sec:num_eval}

The performance of the OSE is evaluated next. Precisely,
\begin{itemize}
    \item the state $\bx$ has dimension $N=15$. The state variation $\bdelta$ is drawn from a uniform distribution $\mathcal U(-\Delta_x/2, \Delta_x/2)$, for every~$t$, with $\Delta_x = 1$;  
    \item the measurement vector $\by$ has, for simplicity, fixed dimension $M_t = M = 3$;
    \item at every time step $t$, the model matrix $\bA(t)$ is chosen from a library of 10 standard normal random variables matrices of dimension $M$ by $N$. Each matrix is then scaled so that Frobenius norm is equal to 1. 
\end{itemize}
Five thousands Monte Carlo simulations are run in both the bounded noise and the stochastic noise case. In every simulation run, the sequence of model matrices $\{\mathbf{A}(t)\}_{t\geq1}$ is generated by randomly selecting a matrix from the matrices library so that $\tau = 4$.
Since the same library of model matrices is used, all the Monte Carlo simulations share the same values of $c$, $\tau$ and $\overline{\lambda}$.

\begin{figure}[t]
    \centering
    \includegraphics[width=0.45\textwidth]{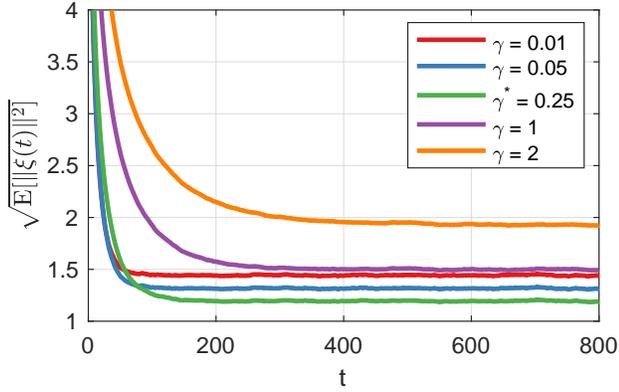}
    \caption{Mean estimation error vs. time under different inertia parameter $\gamma$ settings, averaged over 5,000 Monte Carlo simulations under the bounded noise case.}
    \label{fig:xerr_vs_t_bounded_WLS-guido-001_WLS-guido-005_WLS-guido-025_WLS-guido-1_WLS-guido-2}
\end{figure}

\begin{figure}[t]
\centering
\includegraphics[width=0.4\textwidth]{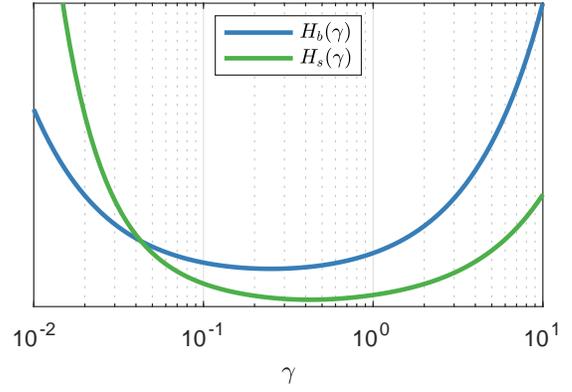}
\caption{Shape of the bounds $H_b(\gamma)$ and $H_s(\gamma)$.}
\label{fig:bound}
\end{figure}

\paragraph{Bounded Noise Case}
Here, the noise vector $\bn(t)$ is generated by drawing from a uniform distribution $\mathcal U(-\Delta_n/2, \Delta_n/2)$, with $\Delta_n = 1$.
For every $t$, matrix $\bQ_t$ is set to $\bQ_t = \bI$. 
Figure~\ref{fig:xerr_vs_t_bounded_WLS-guido-001_WLS-guido-005_WLS-guido-025_WLS-guido-1_WLS-guido-2} reports the average distance between the actual estimate and the true state for different choices of $\gamma$.
According to equation~\eqref{eq:gamm_star_bnd}, the best inertia parameter should be $\gamma^*=0.25$.
Figure~\ref{fig:bound} reports the shape of the bound~\eqref{eq:_xi_bnd_bnd}, as a function of the inertia parameter $\gamma$ and denoted as 
$$H_b(\gamma):=\tau \Big(\Delta_x + \frac 1 \gamma c \Delta_n\Big) \Big(1+ \frac{\gamma}{\bar \lambda}\Big).$$
%

\begin{figure}[t]
    \centering
    \includegraphics[width=0.45\textwidth]{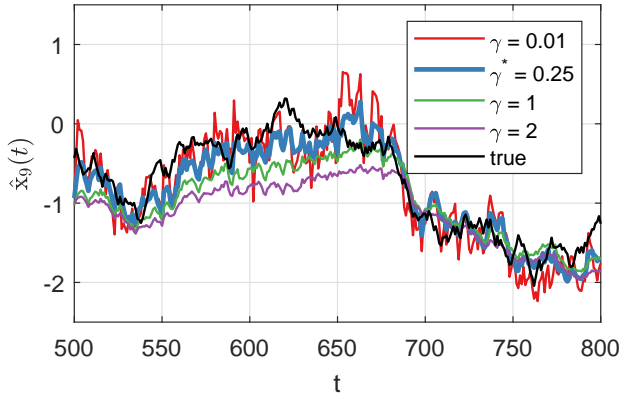}
    \caption{Estimation of $x_9(t)$ over time under different inertia parameter settings with bounded noise.}
    \label{fig:tracking_sample_bounded}
\end{figure}

If the inertia parameter is chosen too small, e.g. see $\gamma=0.01$, the estimation uses almost no past information and becomes very sensitive to the noise. On the other hand, if the inertia parameter is too large, e.g. see $\gamma = 2$, the estimate does not track promptly the true state. When the inertia parameter strikes a balance between new and old information, it can track the true value relatively closely.
Finally, Figure \ref{fig:tracking_sample_bounded} shows the estimation of one particular state element over time in one particular Monte Carlo run. The curve associated with $\gamma^*$ is the best in tracking the true state trajectory.

\paragraph{Stochastic Noise Case}
\label{sub:bnd_noise}

\begin{figure}[t]
    \centering
    \includegraphics[width=0.45\textwidth]{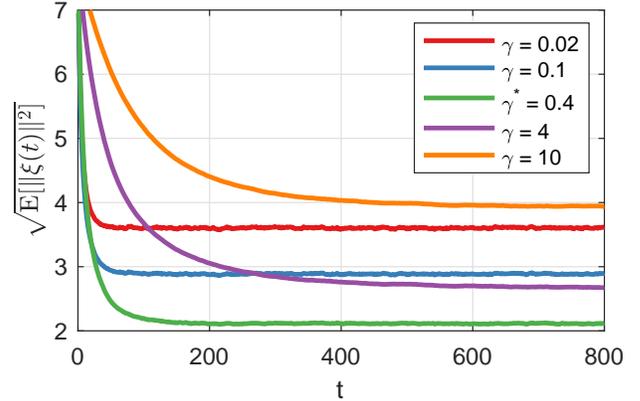}
    \caption{Root mean squared estimation error vs. time under different inertia parameter $\gamma$ settings, averaged over 5,000 Monte Carlo simulations under the stochastic noise case.}
    \label{fig:xerr_vs_t_stochastic_WLS-guido-002_WLS-guido-01_WLS-guido-04_WLS-guido-4_WLS-guido-10}
\end{figure}

Here, the noise vector $\bn(t)$ is generated by drawing from a Gaussian distribution with zero mean and diagonal finite covariance, namely, $\bn(t) \sim \mathcal N(\0,\Delta_n \bI)$ for every $t$, with $\Delta_n = 0.25$.
For every $t$, matrix $\bQ_t$ is set to $\bQ_t = \Delta_n \bI$. The average distance between the actual estimate and the true state for different choices of $\gamma$ is shown in Figure~\ref{fig:xerr_vs_t_stochastic_WLS-guido-002_WLS-guido-01_WLS-guido-04_WLS-guido-4_WLS-guido-10}.
Minimizing the upper bound provided in~\eqref{eq:ave_xi_stoch}, whose shape is in Figure~\ref{fig:bound} and that is denoted as
$$H_s(\gamma):=\tau  \sqrt{\frac{\bar m^2}{\gamma^4} + \Delta_x^2} \left(1 + \frac{\gamma}{\bar \lambda} \right),$$
yields to the theoretical optimal inertia parameter $\gamma^*=0.4$.
The inertia parameter, $\gamma^*=0.4$ gives the experimental low average-case error. Increasing the inertia parameter to $\gamma=25\gamma^*=10$ or decreasing the intertia parameter to $\gamma= \frac{1}{20}\gamma^*=0.02$ almost doubles the error.

\begin{figure}[t]
    \centering
    \includegraphics[width=0.45\textwidth]{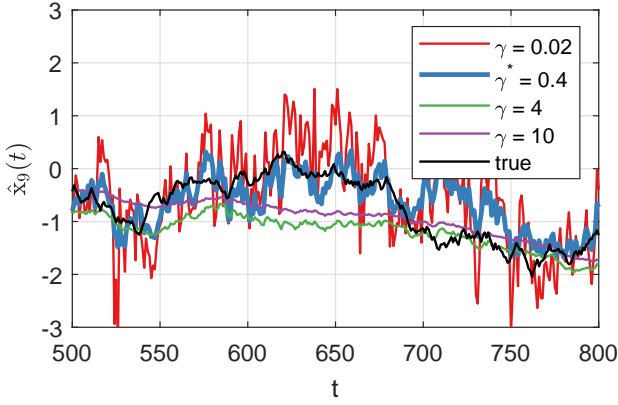}
    \caption{Estimation of $x_9(t)$ over time under different inertia parameter settings with stochastic noise.}
    \label{fig:tracking_sample_stochastic}
\end{figure}

Finally, looking at the estimate of one particular state over time in Figure~\ref{fig:tracking_sample_stochastic}, the results are very similar to that of the bounded noise; an inertia parameter that is too large lags and does not respond immediately to changes in the state, and an inertia parameter that is too small is very sensitive to the noise.
However, the inertia parameter $\gamma^*$ that minimizes the worst-case error strikes a balance between the two extremes.

\section{Conclusion}\label{sec:conclusions}
We have proposed a dynamic state estimation algorithm for linear time-varying systems. The estimator has a recursive expression in which the new estimate is found as a function of the previous estimate and of the gathered measurements. The estimator is designed to tackle  the cases in which the system is not observable, namely, when the measurements do not contain enough information to reconstruct the entire system state. The estimation error 
was proved to be bounded under mild assumptions. Future research directions include considering a non-linear measurement model instead of~\eqref{eq:sys_model}; and analyzing the case in which measurements may suffer from random transmitting delays.
\appendix


%

\begin{IEEEproof}[Proof of Proposition~\ref{prop:asymp_stable}]
Firstly, heed that equation~\eqref{eq:svd_Lam} implies for every $t$ that $\| \bLambda(t) \| \leq 1$
yielding $\Big \| \prod_{k=0}^{\tau} \bLambda(t + k) \Big \| \leq 1$.
Now consider a vector $\bx$, $\|\bx \| = 1$, and assume that
\begin{equation}
    \Big \| \prod_{k=0}^{\tau} \bLambda(t + k) \bx \Big \| = 1
\label{eq:norm_prod_eq1}
\end{equation}
For equation~\eqref{eq:norm_prod_eq1} to hold, it must be that $\| \bLambda(t) \bx \| = 1$, $\| \bLambda(t+1) \bLambda(t) \bx \| = 1$ , $\| \bLambda(t+2) \bLambda(t+1) \bLambda(t) \bx \| = 1$, 
and so on.
Consider now the decomposition of $\bLambda(t)$, given by~\eqref{eq:svd_Lam}. Since $\begin{bmatrix}
\bU(t) & \bV(t)]
\end{bmatrix}$ spans $\mathbb R^N$, the vector $\bx$ can be written as
\begin{equation}
\bx = 
\begin{bmatrix}
\bU(t) & \bV(t)
\end{bmatrix}
\begin{bmatrix}
\bbeta_u(t) \\ \bbeta_v(t)
\end{bmatrix}
\label{eq:v_dec}
\end{equation}
The product $ \bLambda(t) \bx$ can expressed as

\begin{small}
\begin{align*}
& \begin{bmatrix}
\bU(t)^\top \\ \bV(t)^\top
\end{bmatrix}^\top
\begin{bmatrix}
\diag \left ( \left\{ \frac{\gamma}{\gamma + \lambda_i(t)} \right\} \right)& \0 \\
\0 & \bI
\end{bmatrix}
\begin{bmatrix}
\bU^\top(t) \\ \bV^\top(t)
\end{bmatrix}
\begin{bmatrix}
\bU(t)^\top \\ \bV(t)^\top
\end{bmatrix}^\top
\begin{bmatrix}
\bbeta_u(t) \\ \bbeta_v(t)
\end{bmatrix}\\
& = \begin{bmatrix}
\bU(t)& \bV(t)
\end{bmatrix}
\begin{bmatrix}
\diag \left ( \left\{ \frac{\gamma}{\gamma + \lambda_i(t)} \right\} \right) \bbeta_u(t)\\
 \bbeta_v(t)
\end{bmatrix}
\end{align*}
\end{small}

%

It is then easy to see that $\| \bLambda(t) \bx\| = 1$ if and only if $\bx = \bV(t) \bbeta_v(t)$. Moreover, in this case $ \bLambda(t) \bx = \bx$, i.e., $\bx$ is an eigenvector of $ \bLambda(t)$ associated with the eigenvalue 1 and $\bx \in \ker \bA(t)$.
Hence, it holds
\begin{equation}
\bLambda(t+1) \bLambda(t) \bx = \bLambda(t+1) \bx    
\end{equation}
Again, $\|\bLambda(t+1) \bx\| = 1$ if and only if $\bx$ is an eigenvector of $ \bLambda(t+1)$ associated with the eigenvalue 1 and $\bx \in \ker \bA(t+1)$.

By iterating the previous reasoning eventually we can state that equation~\eqref{eq:norm_prod_eq1} holds only if $\bx \in \ker \bA(t),  \ker \bA(t+1), \dots,  \ker \bA(t+\tau) $ contradicting  Assumption~\ref{ass:tau}. Thus,
\begin{align*}
    \Big \| \prod_{k=0}^{\tau-1} \bLambda(t + k) & \bx \Big \| \leq \max_{0\leq k \leq \tau-1} \left\{\frac{\gamma}{\gamma + \lambda_1(k)} \right \}  \\
    & \leq \max_{t} \left\{\frac{\gamma}{\gamma + \lambda_i(t)} \right \} = \psi < 1 
\end{align*}
The asymptotic stability of~\eqref{eq:din_sys} follows directly from~\eqref{eq:norm_prod_str_low1}.
\end{IEEEproof}


\begin{IEEEproof}[Proof of Theorem~\ref{thm:err_bounded}]
Taking the norm on both sides of~\eqref{eq:xi(T)} and using the triangle inequality yields
\begin{align}
    & \|\bxi(T)\| \leq  \left\|\prod_{t=1}^{T} \bLambda(t) \bxi(0) \right\| \nonumber \\
    & \hspace{.5cm} + \sum_{t=1}^T  \left\|\prod_{k=t}^T \bLambda(k) \left(\frac 1 \gamma \bA(t)^\top \bQ_t^{-1} \bn(t) -\bdelta(t) \right) \right\|
    \nonumber \\
    & \leq \left\|\prod_{t=1}^{T} \bLambda(i) \right\| \left\| \bxi(0) \right\| \nonumber \\
    &  \hspace{.5cm} + \sum_{t=1}^T  \left\|\prod_{k=t}^T \bLambda(k) \right\| \left\| \left(\frac 1 \gamma \bA(t)^\top \bQ_t^{-1} \bn(t) -\bdelta(t) \right) \right\|.
    \label{eq:xi(T)_2}
\end{align}
The norm of $\frac 1 \gamma \bA(t)^\top \bQ_t^{-1} \bn(t) -\bdelta(t)$ in the second terms on the right-hand-side of~\eqref{eq:xi(T)_2} can be bounded as: 
\begin{equation*}
\left\| \left(\frac 1 \gamma \bA(t)^\top \bQ_t^{-1} \bn(t) -\bdelta(t) \right) \right\| \leq  \Delta_{x}(t) + \frac{c_t}{\gamma} \Delta_{n}(t)  \, .
\end{equation*}
Let $\rho$ and $r$ be scalars such that, for any $t'$, $t' = \rho \tau + r$, with $r < \tau$, namely, $\rho = \left \lfloor{\frac{t'}{\tau}}\right \rfloor$. Then, it holds
\begin{align*}
\left\| \prod_{i=1}^{t^\prime} \bLambda (i)  \right\|& \leq \left\|\prod_{i=1}^{\tau} \bLambda(i)  \right\| \left\|\prod_{i=\tau+1}^{2\tau} \bLambda(i)  \right\| \cdot \cdot \cdot \nonumber \\
&  \left\| \prod_{i=(\rho-1)\tau+1}^{\rho \tau} \bLambda(i)  \right\|  \left\| \prod_{i=\rho\tau+1}^{t^\prime} \bLambda(i)  \right\| \leq 
\underbrace{\psi \psi  \cdot \cdot \cdot \psi}_{\rho~\textrm{times}} \cdot 1
\end{align*}
where the last step is because of Proposition~\ref{prop:asymp_stable}. Equation~\eqref{eq:_xi_bnd_bnd_T} then follows. 

To prove equation~\eqref{eq:_xi_bnd_bnd}, express $T$ as $T = \rho \tau + r$, with $r < \tau$. Note that, as $T$ goes to infinity, equation~\eqref{eq:xi(T)_2} tends to 
\begin{align*}
& \sum_{\phi = 0}^{\rho} \sum_{t= \phi \tau + 1}^{\min \{(\phi +1)\tau, T \}}  \left \| \prod_{k=t}^T  \bLambda(k)  \right \| \left \| \frac 1 \gamma \bA(t)^\top \bQ_t^{-1} \bn(t) -\bdelta(t) \right \| \leq  \\
&\Big(\frac 1 \gamma c \Delta_n + \Delta_x \Big) 
\sum_{\phi = 0}^{\rho}  \sum_{t= \phi \tau + 1}^{\min \{(\phi +1)\tau, T \}} \left \| \prod_{k=t}^T \bLambda(k)  \right \| \leq  \\
&\Big(\frac 1 \gamma c \Delta_n + \Delta_x \Big) \sum_{\phi = 0}^{\rho} \tau \psi^{\phi} \leq \tau \Big(\frac 1 \gamma c \Delta_n + \Delta_x \Big) \sum_{\phi = 0}^{\infty} \psi^{\phi} \leq  \\
& = \tau \Big(\Delta_x + \frac 1 \gamma c \Delta_n\Big) \Big(1+ \frac{\gamma}{\bar \lambda}\Big) .
\end{align*}
since the first term of~\eqref{eq:xi(T)_2} vanishes due to Proposition~\ref{prop:asymp_stable} and 
\begin{align*}
& \left \| \frac 1 \gamma \bA(t)^\top \bQ_t^{-1} \bn(t) -\bdelta(t)  \right \| \leq  \left \| \bdelta(t)  \right \| + \left \| \frac 1 \gamma \bA(t)^\top \bQ_t^{-1} \bn(t) \right \| \\
&\leq \quad \Delta_x + \frac{\Delta_n}{\gamma} \max_t \{ \| \bA(t)^\top \bQ_t^{-1} \| \} = \Delta_x + \frac 1 \gamma c \Delta_n .
\end{align*}
Finally, equation~\eqref{eq:gamm_star_bnd} can be easily found by minimizing the left hand side of~\eqref{eq:_xi_bnd_bnd}.
\end{IEEEproof}


\begin{IEEEproof}[Proof of Proposition~\ref{prop:sigma_asymp_stable}]
First, we characterize the eigenvalues and the eigenvectors of $\bF(t)$.
Let $\bx_i$ and $\bx_j$ be two eigenvectors of $\bLambda(t)$ associated with two eigenvalues $\mu_i$ and $\mu_j$, i.e., $\bLambda(t) \bx_i = \mu_i, \bLambda(t) \bx_j = \mu_j$. Then, $\bx_i \otimes \bx_j$ is an eigenvector of $\bF(t)$ associated with the eigenvalue $\mu_i \mu_j$, since
\begin{align}
\bF(t) (\bx_i  & \otimes \bx_j) = (\bLambda(t) \otimes \bLambda(t)) (\bx_i \otimes \bx_j) \notag \\
& = (\bLambda(t) \bx_i) \otimes (\bLambda(t) \bx_j) = \mu_i \mu_j \bx_i \otimes \bx_j. 
\label{eq:egvct_F}
\end{align}
Hence, the spectrum of $\bF(t)$ is given by
\begin{align}
& \text{eig } \bF(t) = \Big \{1, \frac{\gamma}{\gamma + \lambda_1(t)}, \dots, \frac{\gamma}{\gamma + \lambda_{I_t}(t)}, \notag \\
&\frac{\gamma}{\gamma + \lambda_{1}(t)} \frac{\gamma}{\gamma + \lambda_{2}(t)}, \dots,\frac{\gamma}{\gamma + \lambda_{i}(t)} \frac{\gamma}{\gamma + \lambda_{j}(t)},\dots \Big\}.    
\end{align}
where $1$ has multiplicity $k_t^2$, each $\frac{\gamma}{\gamma + \lambda_i}$ has multiplicity $2 I_t$ and each $\frac{\gamma}{\gamma + \lambda_{i}(t)} \frac{\gamma}{\gamma + \lambda_{j}(t)}$ has multiplicity 1. Heed that the biggest eigenvalue of $\bF(t)$ smaller than 1, similarly to $\bLambda(t)$, is $\frac{\gamma}{\gamma + \lambda_1}$. From~\eqref{eq:egvct_F} it is also clear that the eigenvectors of $\bF(t)$ associated with the eigenvalue 1 have the form $\bv_i \otimes \bv_j$, where $\bv_i, \bv_j$ are the $i$-th and the $j$-th column of $\bV(t)$, respectively.
Now consider any vector $\bx \in \mathbb{R}^{N^2}$, with $\|\bx\| = 1$ and the product  $\left \| \prod_{k = 0}^{\tau} \bF(t + k) \right \|$. Retracing the same reasoning used in the proof of Proposition~\ref{prop:asymp_stable}, it can be shown that $\left \| \prod_{k = 0}^{\tau} \bF(t + k) \bx  \right\| = 1$ if and only if $\bx \in \ker \bF(t+k) = \text{span } \{ \bv_i \otimes \bv_j, \bv_i, \bv_j \in \ker \bA(t+k) \}$, for every $k = 0,\dots, \tau$, contradicting Assumption~\ref{ass:tau}.
Hence,
\begin{align*}
    \Big \| \prod_{k=0}^{\tau-1} \bF(t + k) & \bx \Big \| \leq \max_{0\leq k \leq \tau} \left\{\frac{\gamma}{\gamma + \lambda_1(t+k)} \right \}  \leq \psi < 1 
\end{align*}
The asymptotic stability of~\eqref{eq:F_din_sys} follows directly from~\eqref{eq:F_norm_prod_str_low1}.
\end{IEEEproof}

\begin{IEEEproof}[Proof of Theorem~\ref{thm:err_stoch}]
Applying the triangle inequality to equations~\eqref{eq:mu(T)} and~\eqref{eq:sigma(T)}, the norm of $\bmu(T)$ and $\bsigma(T)$ can be upper bounded by
\begin{small}
\begin{align}
&\|\bmu(T)\| \leq \Big \|\prod_{t=1}^{T} \bLambda(t) \bxi(0) \Big \| + \Big \|\sum_{t=1}^T \prod_{k=t}^T \bLambda(k)  \bdelta(t) \Big \| \label{eq:norm_mu(T)}\\
&\| \bsigma(T) \| \leq \left\| \prod_{t=1}^T \bF(t) \bsigma(0) \right \|  +  \left\| \frac{1}{\gamma^2} \sum_{t=1}^{T} \prod_{k=t}^T \bF(k) \bC(t) \bm(t) \right \|
\label{eq:norm_sigma(T)}.
\end{align}
\end{small}
Note that $\| \delta(t) \| \leq \Delta_x(t)$ and that $\|\bC(t) \bm(t) \| \leq C(t)  m(t)$. Equations~\eqref{eq:mu_stoch_T} and \eqref{eq:sigma_stoch_T} can be obtained by retracing the same steps used to prove~\eqref{eq:_xi_bnd_bnd_T}. 

The first term in the RHS of~\eqref{eq:norm_mu(T)} tends to zero as $T$ goes to infinity, due to Proposition~\ref{prop:asymp_stable}. Consider now the second term and let $\rho$ and $r$ be scalars such that $T = \rho \tau + r$, with $r < \tau$. It holds
\begin{align*}
&\Big \|\sum_{t=1}^T \prod_{k=t}^T \bLambda(k)  \bdelta(t) \Big \| =
 \Big \| \sum_{\phi=0}^{\rho} \sum_{t=\phi \tau +1 }^{\min\{(\phi + 1) \tau, T \}} \prod_{k=t}^T \bLambda(k) \bdelta(t) \Big \| \\
& \quad \leq  \Delta_x \sum_{\phi=0}^{\rho} \sum_{t=\phi \tau +1 }^{\min\{(\phi + 1) \tau, T \}} \Big \| \prod_{k=t}^T \bLambda(k) \Big \|  \leq  \Delta_x \sum_{\phi=0}^{\rho} \tau \psi^{\phi}  \\
& \quad \leq  \Delta_x \tau \sum_{\phi=0}^{\infty}  \psi^{\phi} \leq \Delta_x \tau \left(1 + \frac{\gamma}{\bar \lambda} \right).
\end{align*}
Similarly, as $T$ goes to infinity, Proposition~\ref{prop:sigma_asymp_stable} ensures that the first term of the right hand side of~\eqref{eq:norm_sigma(T)} goes to zero. Now consider the second term. Then, we have

\begin{align*}
\Bigg\|  \sum_{t=1}^{T} \prod_{k=t}^T & \frac{\bF(k)}{\gamma^2} \bC(t) \bm(t) \Bigg \| =  \\
& = \left\| \sum_{\phi=0}^{\rho} \sum_{t=\phi \tau +1 }^{\min\{(\phi + 1) \tau, T \}} \prod_{k=t}^T \frac{\bF(k)}{\gamma^2}  \bC(t)  \bm(t) \right \| \\
& \leq \frac{C m}{\gamma^2} \sum_{\phi=0}^{\rho} \sum_{t=\phi \tau +1 }^{\min\{(\phi + 1) \tau, T \}} \left\| \prod_{k=t}^T \bF(k) \right \|  \leq \frac{\bar m}{\gamma^2} \sum_{\phi=0}^{\rho} \tau \psi^\phi \\
& \leq \frac{ m \tau}{\gamma^2} \sum_{\phi=0}^{\infty} \psi^\phi = \frac{ m \tau}{\gamma^2} \left(1 + \frac{\gamma}{\bar \lambda} \right).
\end{align*}

Finally, to prove equation~\eqref{eq:ave_xi_stoch}, heed that
\begin{align*}
\mathbb{E}[\bxi^\top(t) & \bxi(t)] = \mathbb{E}[(\bxi(t) - \bmu(t))^\top (\bxi(t) - \bmu(t))] + \bmu(t)^\top \bmu(t) \\     
& = \mathbb{E}[\trace((\bxi(t) - \bmu(t))(\bxi(t) - \bmu(t))^\top) ]  + \bmu(t)^\top \bmu(t) \\
& = \|\bSigma(t)\|_F^2 + \| \bmu(t)\|^2 \\
& = \tau^2 \left(1 + \frac{\gamma}{\bar \lambda} \right)^2  \left(\frac{C^2 m^2}{\gamma^4} + \Delta_x^2 \right)
\end{align*}
where we used equation~\eqref{eq:mu_stoch} and \eqref{eq:sigma_stoch}.
\end{IEEEproof}

\bibliographystyle{IEEEtran}
\bibliography{myabrv,power}

\end{document}